# Strength-dependent Transition of Graphite Under Shock Condition Resolved by First Principles

Gu-Wen Chen[1], Liang Xu[2], Yao-Ming Li[3], Zhi-Pan Liu[4]*, Sheng-Cai Zhu[1]*

1. School of Materials, Shenzhen Campus of Sun Yat-sen University, Shenzhen 518107, China.

2. National Key Laboratory of Shock Wave and Detonation Physics, Institute of Fluid Physics, China Academy of Engineering Physics, Mianyang 621900, China.

3. College of Aerospace Engineering, Nanjing University of Aeronautics and Astronautics, Nanjing 210016, China

4. Collaborative Innovation Center of Chemistry for Energy Material, Shanghai Key Laboratory of Molecular Catalysis and Innovative Materials, Key Laboratory of Computational Physical Science, Department of Chemistry, Fudan University, Shanghai 200433, China.

*Corresponding author. Email: S.Z(zhushc@mail.sysu.edu.cn), Z.L(zpliu@fudan.edu.cn)

**Abstract**

The shock strength dependent formation of diamond represents one of the most intriguing questions in graphite research. Using *ab initio* DFT-trained carbon GNN model, we observe a strength-dependent graphite transition under shock. The poor sliding caused by scarce sliding time under high-strength shock forms hexagonal diamond with an orientation of $(001)_G//(100)_{HD}+[010]_G//[010]_{HD}$; under low-strength shock, cubic diamond forms after enough sliding time, unveiling the strength-dependent graphite transition. We provide computational evidence of the strength-dependent graphite transition from first principles, clarifying the long-term shock-induced hexagonal formation and structural strength-dependent trend source.

According to the strain rate of compression, pressure loading can be divided into static (or quasistatic) pressure loading and shock loading. Apart from providing a fast path to reach high-pressure condition, the nonequilibrium effect, strong shear action and obvious plastic deformation in dynamic loading, shock is of significant scientific and engineering interest [1,2]. For nonequilibrium compression, shock compression might be fundamentally different from static compression. To date, dynamic compression is a general route for synthesizing novel materials. Recent advancements in experimental techniques have enabled us to directly probe phase transitions and their kinetics at the atomic scale. Time-resolved X-ray diffraction, in particular, has been coupled with shock compression drivers in recent years [3–5], offering a detailed and precise view of the underlying processes of shock compression. Thus far, the understanding of microstructure and phase transition coupling during shock, which significantly influences phase transition kinetics, is beyond the scope of state-of-the-art experiments. Thus, it is imperative to provide mechanistic insights through simulations to achieve a profound understanding of the phase transition during shock.

Carbon is a highly pivotal element, and the phase transition from graphite (G) to diamond is regarded as one of the most classic phase transitions. This phase transition has been extensively studied [6–10], is not yet understood. Cubic diamond (CD) and hexagonal diamond (HD; also known as lonsdaleite) are the most commonly reported high-pressure carbon phases. Hexagonal diamond is of significant interest due to its excellent elastic and mechanical properties [11,12], making it potentially useful for various applications. However, hexagonal diamond has only been discovered during the impact of meteorites on Earth [13], suggesting that it can be quenched to ambient conditions as a metastable phase; moreover, the shock-induced G-to-HD transition has been reproduced in the laboratory by *in situ* X-ray diffraction (XRD) [14]. The shock-induced graphite-to-diamond transition is proven to be strength dependent. Under low-strength shock conditions (such as detonation conditions [15–17]), the main product is thermally stable CD in the recovered materials. Recently, under high-strength shock conditions, Kraus et al. [18] and Turneaure et al. [14] found that highly oriented pyrolytic graphite transforms into metastable HD. While these efforts provide strong evidence consistent with the shock-induced formation of HD, the critical shock strength leading to HD has not been confirmed due to the difficulty of its identification in experiments. Doubts have been raised upon HD identification in various diamond-like materials [19–21]. Kraus et al. [18] reported that only forms above 170-GPa shock for shocked pyrolytic graphite, while Turneaure et al. [14] reported that HD only forms under ~50-GPa shock. In this context, definitively resolving the strength-dependent structure evolution process under shock compression where phase transition occurs and further illuminating the original phase selectivity are efforts of practical and fundamental importance.

Large-scale computational simulations, as an important step toward understanding the microscopic nature of the phase transition selectivity of graphite during shock, have been conducted extensively in recent years to follow the strength-dependent transition in real time [22–28]. For example, via *ab initio* molecular dynamics simulation, Mundy et al. [27] found that graphite transforms to an intermediate layered diamond before it transforms into cubic diamond under shock. While integrating classical adaptive intermolecular reactive empirical bond order–Morse (AIREBO-M) [28] and long-range carbon bond-order potential II (LCBOPII) [26] potentials, Sun and Pineau found that graphite only yields CD as the main product regardless of the shock strength, with HD appearing as a twin boundary; this result is inconsistent with the experimental findings. Unfortunately, to date, neither *ab initio* [27] nor empirical force field simulations [26,28] can reproduce the experimental shock-induced HD formation [14,18]. The *ab initio* simulations cannot reproduce the anisotropy of the graphite system due to the limited system size and high computational cost, while classical force fields are overly simple in functional forms; thus, they cannot capture the interatomic interactions of carbon since they are parametrized to match limited experimental thermophysical data and fail to reach the requisite accuracy. Even a more complex angular-dependent-potential (ADP) [29], which was parametrized recently based on *ab initio* calculations by extending the multipole expansion approach to the quadrupole term, still fail to reproduce pure HD formation under high-strength shock (Fig. S5 in the supporting information). Thus far, no conclusive computational evidence of shock strength-dependent formation of pure HD is available, and an effective method that can compromise between accuracy and speed is needed to discover the shock-induced HD formation mechanism.

Recently, machine-learning (ML)-based atomic simulations have emerged as a major step forward for accelerating material research, which relies on predictive surrogate ML models, such as artificial neural networks (ANNs), to evaluate the atomic energy and force of complex potential energy surfaces (PESs). The neural network (NN) models can meet the needed level of accuracy and properly describe chemical reactions by pretraining a large representative PES dataset calculated from quantum mechanics, from which the NN models capture the many-body correlations and complex polarizability effects of multidimensional PESs. The method pushes the boundaries of atomic simulations to a vast array of complex material and chemical reaction systems [30–32], showing great advantages in representativity, extensity, and continuity [33] relative to classical force field methods. In this study, by using an *ab initio*-based global neural network (GNN) model of carbon trained on a density functional theory (DFT)-based PES dataset globally sampled from stochastic surface walking (SSW) simulations [34], we performed unbiased molecular dynamics (MD) simulations to study the strength-dependent graphite-to-diamond phase transition mechanism under shock condition. With the high accuracy and high speed of the GNN potential, we could scan the phase spaces of carbon allotropes, all possible interface structures and phase transition pathways [25], and we can identify the strength-dependent graphite-to-diamond phase transition mechanism. By varying the particle velocities from 1 to 8 km/s, we found that graphite transforms into CD at a speed of 3.0–4.5 km/s, into HD at a speed above 4.5 km/s with an orientation relation of $(001)_G//(100)_{HD}+[010]_G//[010]_{HD}$, and into an amorphous carbon/HD phase at a speed above 6 km/s. Our simulations are uniquely consistent with the experimental shock-based formation of HD [14] and reveal the microstructural origin of the strength-dependent transition in graphite shock experiments. The proposed transition mechanism provides a basis for HD identification and therefore helps to further determine the critical shock strength to HD.

The details of the GNN potential can be found in the Supporting Information (SI). The benchmark for the accuracy of this potential is presented in Figs. S1 and Table S2, showing values comparable to those of DFT calculations. The shock MD simulations are performed in the large-scale atomic/molecular massively parallel simulator (LAMMPS) package [35], and a detailed simulation method is presented in the SI. To rule out possible sizing effects caused by the periodic boundary conditions, we constructed a series of models with different sizes for shock simulation containing 1040, 4480, 7200, 9720, 10560, and 13728 atoms. For models with over 9720 atoms, the simulation results show no significant difference, indicating that the system size is sufficiently large to obtain reliable results. We chose the 13728 atom (2.57 nm*2.72 nm in the x-y plane, 50 layers) simulation results for detailed analysis.

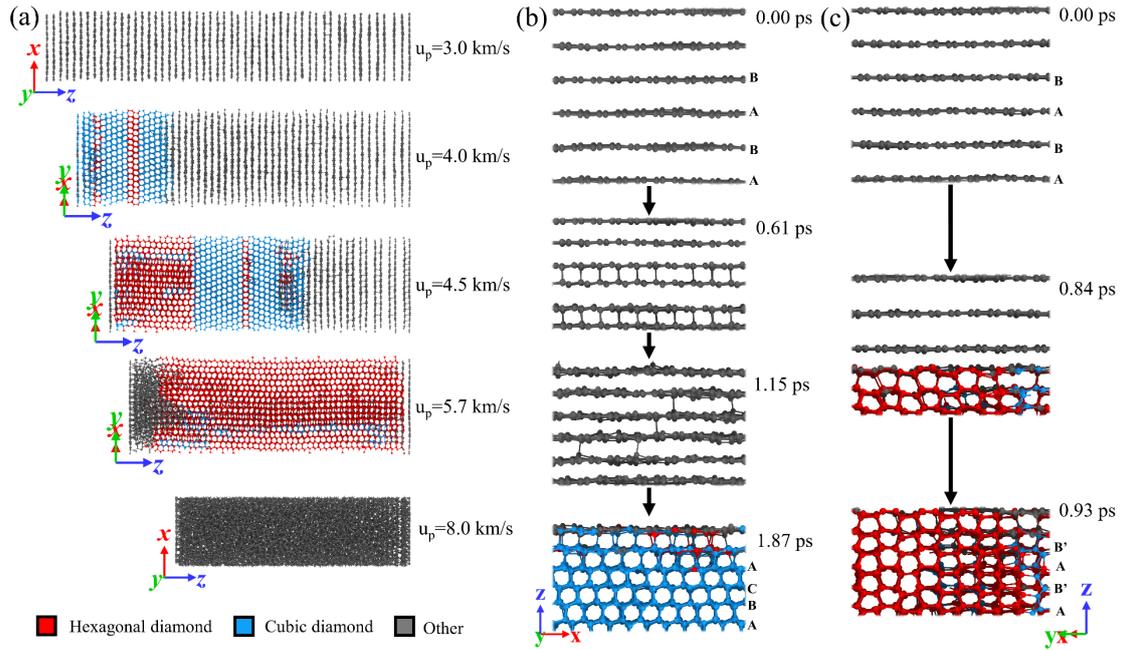

Fig. 1. (a) Partial simulation results with different particle velocities of 3.0, 4.0, 4.5, 5.7, and 7.0 km/s. (b) G/CD transition mechanism seen through the local structural evolution of layers 12–17 in the $u_p$=4.0 km/s case. (c) G/HD transition mechanism seen through the local structural evolution of layers 23–28 in the $u_p$=5.7 km/s case. Red atoms represent the hexagonal diamond phase, blue atoms represent the cubic diamond phase, and gray atoms represent either the graphite or amorphous phase.

The shock product structures of different particle velocities ($u_p$) are presented in Fig. 1a. Other shocked structures and the structural evolution findings from simulations are shown in Figs. S2 and S3, respectively. As shown in Fig. 1a, only elastic compression occurs when $u_p$ to < 3.5 km/s, not a transition. The interlayer spacing is compressed but the typical layered structure is retained without interlayer bonding (Fig. S4a). The stacking order of ABAB is intact under shock. When the $u_p$ increases to 3.5 km/s, the G–CD transition is observed, with the first CD layer emerging at ~0.9 ps. The orientation between graphite and CD is $(001)_G//(111)_{CD} + [100]_G//[110]_{CD}$ (Fig. 1b). A twin structure can be found in some cases (Fig. 1a). When the $u_p$ > 4.8 km/s, metastable HD is the main transitional product with an orientation of $(001)_G//(100)_{HD} + [010]_G//[010]_{HD}$ (Fig. 1c), bypassing the thermally activated G–CD pathway. The first HD layer emerges at ~0.3 ps, much shorter than CD. Defects and deformation are found in the HD phase, but the crystal quality improves as the $u_p$ increases until reaching 5.7 km/s. The details of the G–CD and G–HD transition mechanisms are presented in Figs. 1b and 1c and will be discussed in detail in the following section. Interestingly, a mixed CD-HD phase is found at $u_p$=4.5 km/s. When the $u_p$ is 7 km/s, graphite collapses structurally to an amorphous form, bypassing the above two crystal–crystal pathways. Our simulations and the previous experimental results clearly show that the shock-induced transition of graphite is strength dependent. For comparison, we performed similar simulations using other potentials (Fig S5), including adp, edip, lcbop, and tersoff. Except for adp, which yields a small amount of HD at a 4.0 km/s shock that returns to cubic diamond at a 5.7 km/s shock, none of these empirical simulations can strictly reproduce the experimental shock-induced HD formation. Our *ab initio* NN simulations are uniquely consistent with previous experiments [14], showing great promise for carbon-based transition predict.

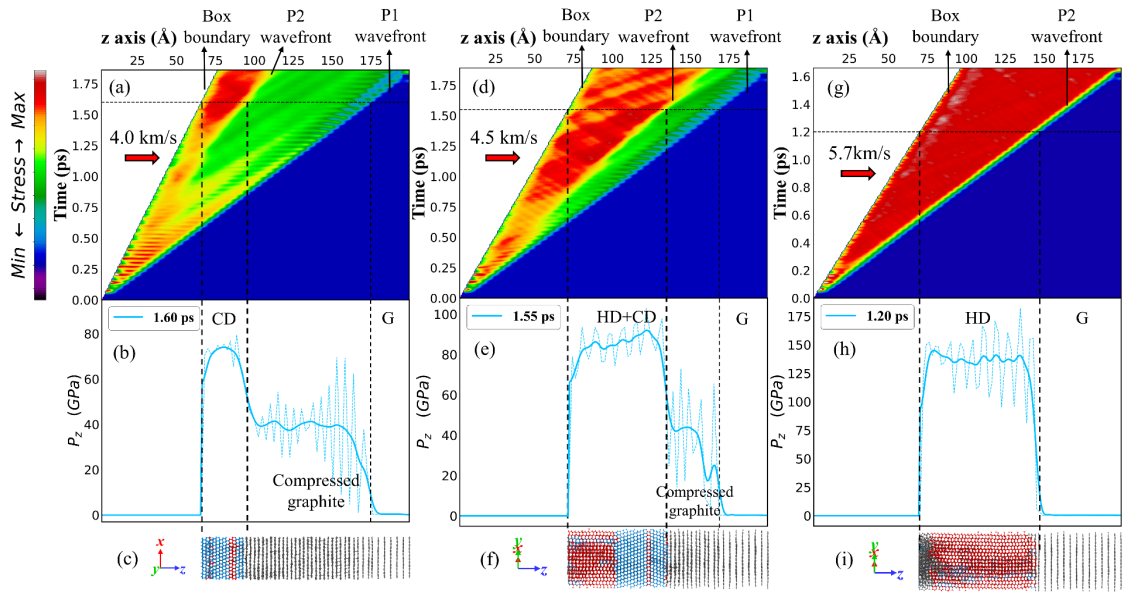

Fig. 2. Layer-averaged longitudinal stress ($P_z$) analysis of three typical cases along the wave propagation direction. (a), (d), (g) Wavefront evolution with time. Red, green, and blue represent high, medium, and low stress, respectively. (b), (e), (h) Wave pattern at specific times depicted by both raw and smoothed data. (c), (f), (i) Structural depiction of the wave pattern at specific times. G: graphite, CD: cubic diamond, and HD: hexagonal diamond.

To clarify the essence of this strength-dependent phenomenon, detailed shock stress ($P_z$) analysis was performed along the wave propagation direction for three typical cases (Fig. 2). The longitudinal stress was evaluated by layer averaging. The wave profiles (in the middle panels) show severe oscillation due to the large interlayer spacing of graphite, which is consistent with the incident wave propagation investigation results by Sun et al. [28]. Here, the wave profiles were filtered smoothly to obtain clear wave patterns. As shown in Fig. 2a–2b, a two-wave pattern is identified in the $u_p$=4.0 km/s case, namely, the elastic wave (P1) and transitional wave (P2). P1 and P2 waves are integrated at first but gradually split into two waves (Fig. 2a) because the P1 wave propagates much faster than the P2 wave. Under the P1 wave, the decrease in interlayer spacing leads to a short-lived layered structure (0.61 ps in Fig. 1b) similar to that proposed by Mundy et al. in their *ab initio* study [27]. In this intermediate state, the graphite layers slide transversely, resulting in a slight stacking variation, but no obvious plane puckering is observed. The intermediate state lasts for ~0.5 ps before reverting to a graphite structure with few interlayer bonds left (1.15 ps in Fig. 1b). When the P2 wave arrives, the graphite plane is bonded with the adjacent CD layer and puckers into a chair conformation. The graphite layer in the G/CD interface continues sliding to fit in the ABCA stacking and finally becomes a new CD layer (1.87 ps in Fig. 1b). The two-wave pattern accompanying CD formation is consistent with previous experimental results [35,36]. In this scenario, the box is divided into three regions: uncompressed graphite, compressed graphite, and diamond (Fig. 2c). Notably, there are some HD layers within the final CD matrix, resulting in stacking faults and twin structures. Fig. 2d–2f show the $u_p$=4.5 km/s results, yielding an HD–CD mix product with a two-wave pattern. However, P1 is closer to P2 in this case than in the 4.0 km/s case. HD only forms in the early stage of shock, while CD formation followed HD formation (Fig. 2f) after the separation of P1 and P2. Intriguingly, during the G–HD transition of this case, the layered intermediate structure directly transforms into HD (Fig. S4b) instead of reverting to the graphite structure, as the G–CD transition shows (Fig. 1b and S4c). Fig. 2g–2i shows the $u_p$=5.7 km/s results with pure metastable HD phase formation. The wave profile and wavefront propagation analysis clearly show an overdriven single-wave pattern, consistent with previous experiments by Kraus et al. [18] and Turneaure et al. [14]. Without the observation of the P1 wave, the short-lived layered structure is absent, and the graphite directly transitions to HD (Fig. 1c). At the wavefront of P2 wave, the graphite plane puckers into a boat configuration and is bonded with the adjacent HD layer to form the G/HD interface. Hindered by the interlayer bonds, the graphite layer in the interface slides only for a short distance to fit in the AB'AB' stacking order. In this case, the simulation box is separated into two regions: uncompressed graphite and HD (Fig. 2h and 2i). Notably, the amorphous phase forms in the early stage of the 5.7 km/s case, the amount of which increases with the increasing particle velocity (Fig. S2). The above analysis indicates that the strength-dependent behavior has a strong relationship with the wave pattern. The CD phase forms with an obvious two-wave pattern, while the HD phase forms with an overdriven single-wave pattern.

According to the orientation relation of transitional products with graphite, layer sliding is a necessary stage during the transition from graphite to diamond. From a topological view, the layer sliding distances for G–CD and G–HD are different. For the G–CD transition, a long distance is needed from ABAB stacking to ABCABC stacking, while for the G–HD transition, a short distance is needed from ABAB to AB'AB' stacking. A previous experimental study [18] suggests that sliding of the graphite basal planes and diamond formation can be prevented under high-strength shock conditions. The above structural analysis indicates that the layer sliding of the G–HD transition is hindered under overdriven single-wave shock conditions.

To confirm the above opinion, we analyzed the layer sliding distance (S) values of individual layers in the aforementioned three cases (Fig. 3). Here, only the effective sliding for stacking transformation before transition completion is considered. As Fig. 3a and 3b show, with sufficient sliding time at $u_p$=4.0 km/s, the CD layer sliding distance reaches 1.8 Å. For $u_p$=4.5 km/s, the stress evolution characteristics of the 5$^{th}$ HD layer and 20$^{th}$ CD layer clearly show that the CD layer has a longer P1 wave duration time than the HD layer. The sliding distance for CD layer formation is ~0.86 Å, while that for HD layer formation is ~0.66 Å. In the $u_p$=5.7 km/s case presenting an overdriven single-wave pattern, only the P2 wave is identified (Fig. 3g). The sliding distance of the 20$^{th}$ HD layer is ~0.48 Å (Fig. 3h).

The above analysis proves that sliding is prevented by insufficient sliding time under high-strength shock conditions. Under high-particle velocity shock, the P1 wave duration is short or even absent, hindering CD formation and favoring HD formation. This phenomenon is further verified by the analysis of the layer groups, as shown in Fig. 3c, 3f, and 3i. The CD layers have significantly larger sliding distances than the HD layers overall, which is related to the shock strength. In addition, the twin structure reduces the requisite sliding distance for CD formation, indicating that a twin structure is formed in CD products. To further confirm our arguments, shock simulation on randomly stacked graphite was conducted (Fig. S6), where the sliding advantage of HD formation is absent. This simulation yields CD as the only product, strongly confirming our conclusions.

In Turneaure's experiments [14], the overdriven single-wave pattern was identified at ~50 GPa, and the HD phase was yielded in this pressure range. Therefore, our work supports the experimental results of Turneaure et al. [14] because our results show that the overdriven single-wave pattern implies prevents sliding and favors HD formation. Our simulations reveal that the transverse sliding is the cornerstone of the shock-strength-dependent phenomenon, and the overdriven single-wave pattern may imply the HD formation in the shock experiment. The proposed mechanism provides an identification basis for the wave pattern for HD formation in shock experiments, further determining the critical shock strength for HD formation.

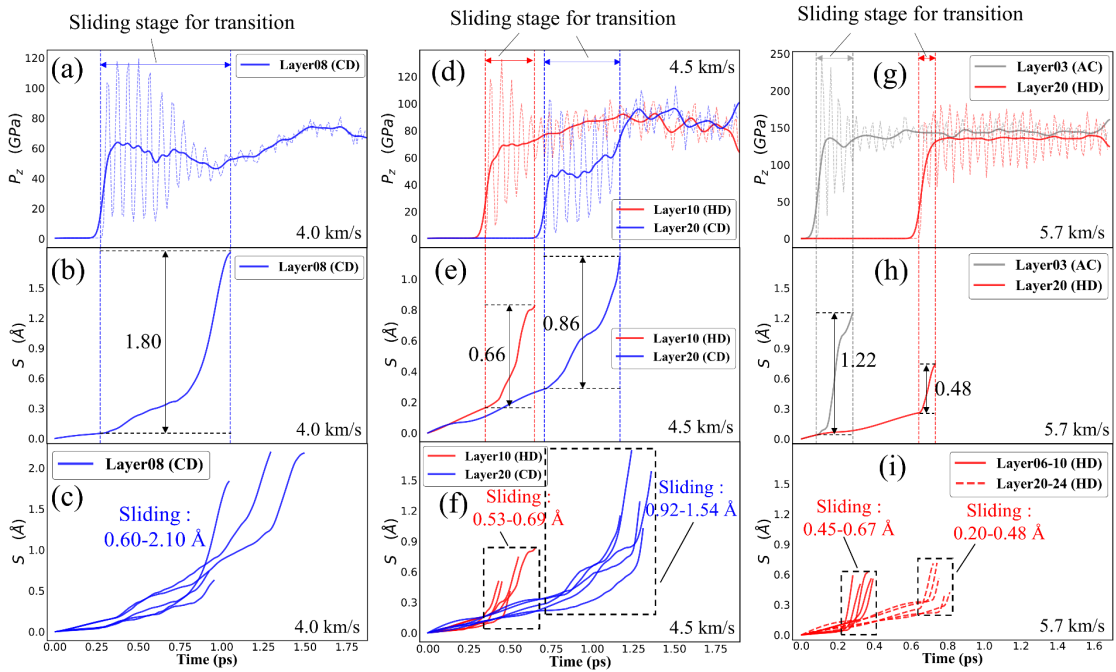

Fig. 3. Shock stress and sliding distance analyses of individual layers for three typical cases: (a–c) $u_p$=4.0 km/s, (d–f) $u_p$=4.5 km/s, and (g–i) $u_p$=5.7 km/s. G: graphite, CD: cubic diamond, HD: hexagonal diamond, and AC: amorphous carbon.

The amount of amorphous carbon phase (AC) increases with increasing particle velocity. The stress and sliding analyses of the $u_p$=5.7 km/s case (Figs. 3g and 3h) show that the sliding time of the AC layer is longer than that of the HD layer, and the peak instantaneous stress of the AC layer is greater than that of the HD layer. Considering the disorder of the amorphous phase, we may not attribute its formation to hindered sliding but to the high instantaneous stress, which induces bond collapse. When the peak instantaneous stress diminishes and becomes stable as the wave propagates, the bonds remain intact, and thus, the diamond structure is maintained, as shown in Fig. S3g. For simulation at high $u_p$ values (eg. $U_p$=8.0 km/s), the peak instantaneous stress is still very high after stabilization; thus, no diamond phase is retained.

In summary, by using an *ab initio* NN model of carbon, our unbiased shock simulations of hexagonal graphite uniquely reproduce the previous experimental results, and reveal the structural origin of the shock-strength-dependent transition. The G/CD transition is observed at particle velocities above 3.5 km/s via a diffusion-free process. The G/HD transition occurs at particle velocities above 4.5 km/s. Evident amorphization is found at particle velocities above 7.0 km/s. The layer sliding distance is the cornerstone of the strength-dependent phenomenon. In low-strength shock cases with a two-wave pattern, sufficient sliding time leads to a long sliding distance, favoring CD formation. Conversely, in high-strength shock cases with an overdriven single-wave pattern, hindered layer sliding leads to a metastable HD phase. Extremely high-strength shock strength may induce bond collapse and consequent amorphization. Our study reveals the long-term unresolved shock-induced HD formation mechanism and provides an identification basis for the wave pattern of this formation in shock experiments. The findings can further determine the critical shock strength for HD formation and elucidates the synthesis of HD. The ML-based *ab initio* approach demonstrated herein can hopefully push the boundaries of computational physics and chemistry for other materials.

## ASSOCIATED CONTENT

**Supporting Information**

The calculation details, components of the NN training dataset, NN potential accuracy benchmarks, shock product structures, structural evolution characteristics of simulations, simulation results with other potentials, results of supplementary simulation, and wave profiles of specific cases can be found in the Supporting Information, which is available free of charge on the website.

**AUTHOR INFORMATIONS**

**CORRESPONDING AUTHOR**

Sheng-Cai Zhu − School of Materials, Shenzhen Campus of Sun Yat-sen University, Shenzhen 518107, China; orcid.org/0000-0003-3311-6723, E-mail: zhushc@mail.sysu.edu.cn

Zhi-Pan Liu- Collaborative Innovation Center of Chemistry for Energy Material, Shanghai Key Laboratory of Molecular Catalysis and Innovative Materials, Key Laboratory of Computational Physical Science, Department of Chemistry, Fudan University, Shanghai 200433, China. E-mail: zpliu@fudan.edu.cn

## AUTHOR CONTRIBUTIONS

# These authors contributed equally: Gu-Wen Chen, Sheng-cai Zhu.

## CONFLICTS OF INTEREST

The authors declare no competing financial interests.

## ACKNOWLEDGEMENTS

This work was supported by the National Science Foundation of China (Grant No: 12274383, 21703004) and Fundamental Research Funds for the Central Universities, Sun Yat-sen University (Grant No. 23qnpy04). We acknowledge the use of computing resources from the Tianhe-2 Supercomputer.

## REFERRENCES